\documentclass[aps,prl,twocolumn,showpacs,floatfix,superscriptaddress]{revtex4}
\usepackage[]{graphicx}

\usepackage{graphicx}
\usepackage{dcolumn}
\usepackage{bm}
\usepackage[usenames]{color}    

\usepackage{siunitx}
\usepackage{enumerate}
\usepackage{lineno}
\usepackage[utf8]{inputenc}
\usepackage[english]{babel}

\begin{document}


\title{Twin beam quantum-enhanced correlated interferometry\\ for testing fundamental physics}

\author{S. T. Pradyumna}
\author{E. Losero}
\affiliation{INRIM, Strada delle Cacce 91, I-10135 Torino, Italy}
\affiliation{Politecnico di Torino, C.so Duca degli Abruzzi, 10100 Torino, Italy}

\author{I. Ruo-Berchera\footnote{\textcolor{black}{e-mail: i.ruoberchera@inrim.it}}}
\author{P. Traina}
\author{M. Zucco}
\affiliation{INRIM, Strada delle Cacce 91, I-10135 Torino, Italy}

\author{C. S. Jacobsen}
\author{U. L. Andersen}
\affiliation{Center for Macroscopic Quantum States (bigQ), Department of Physics, Technical University of Denmark, Fysikvej, 2800 Kgs. Lyngby, Denmark}

\author{I. P. Degiovanni}
\affiliation{INRIM, Strada delle Cacce 91, I-10135 Torino, Italy}

\author{M. Genovese}
\affiliation{INRIM, Strada delle Cacce 91, I-10135 Torino, Italy}
\affiliation{INFN, sezione di Torino, via P. Giuria 1, 10125 Torino, Italy}

\author{T. Gehring}
\affiliation{Center for Macroscopic Quantum States (bigQ), Department of Physics, Technical University of Denmark, Fysikvej, 2800 Kgs. Lyngby, Denmark}

\begin{abstract}
	Quantum metrology deals with improving the resolution of instruments that are otherwise limited by shot-noise and it is therefore a promising avenue for enabling scientific breakthroughs. The advantage can be even more striking when quantum enhancement is combined with correlation techniques among several devices. Here we present and realize a  correlation interferometry scheme exploiting bipartite quantum correlated states injected in two independent interferometers. The scheme outperforms classical ones in detecting a faint signal that may be correlated/uncorrelated between the two devices. \textcolor{black}{We also compare its sensitivity with that obtained for a  pair of two independent squeezed modes, each addressed to one interferometer, for detecting a correlated stochastic signal in the MHz frequency band, being simpler may eventually find application to fundamental physics tests, e.g. searching for the effects predicted by some Planck scale theories.}
	
\end{abstract}
\maketitle

\section*{\textcolor{black}{Introduction}}

Quantum metrology is a sub-field of quantum physics which deals with improving measurement sensitivity beyond the classical limits, by exploiting the properties of quantum systems~\cite{Giovannetti2011, Giovannetti2004, Olivares2018, Adesso2016}. It has already been used to enhance the performance of interferometers~\cite{Caves1981, McKenzie2002,Toth2014, Demkowicz-Dobrzanski2015, Pezze2014, Manceau2017, Boto2000, Huelga1997}, improve phase estimation~\cite{Berni2015} and super-resolution~\cite{DAngelo2001, Oppel2012, Schafermeier2018}, surpass the shot-noise limit in imaging~\cite{Brida2010, Brida2009, Genovese2016} and absorption measurements~\cite{Sabines-Chesterking2017, Losero2018}.  A first level of improvement achieved with quantum light is offered by the use of squeezed states, that has found applications in advanced detection of gravitational waves~\cite{Abadie2011,Aasi2013}. A second level can be obtained by making use of quantum correlations, as entanglement.
For example, in~\cite{Ruo-Berchera2013, Ruo-Berchera2015}, an advantage of using quantum correlated light when comparing signals in a pair of interferometers has been demonstrated. 

Correlation techniques exploiting more than one interferometer are currently employed in fundamental physics, representing the best sensing methodology proposed for a range of predicted fundamental backgrounds  \cite{Romano2017, Akutsu2008, Abbott2017-1, Abbott2017-2,Chou2017, Bassi17} and in particular in the research of  Planck scale effects~\cite{Hogan2012, Hogan17,Chou2016, Chou2017a}. These sources of noise can produce correlated phase fluctuations in two separated interferometers increasing the chance of distinguishing them with respect to other noise sources, thus exceeding the sensitivity of the single device of orders of magnitude. However, in all the experiments reported up to now ~\cite{Chou2016, Chou2017a, Chou2017, Abbott2017-1, Abbott2017-2} the two (or more) interferometers are fed with separated and independent sources of coherent (and in some case squeezed) light, resulting in uncorrelated quantum fluctuations. A double interferometer configuration is for instance the basis of the Fermilab ``holometer''~\cite{Hogan2012, Chou2016}, a device consisting of two co-located 40\,m Michelson interferometers (MIs). The purpose of the holometer is to search for a particular type of correlated background noise, conjectured in some heuristic Planck scale theories and dubbed holographic noise~\cite{Chou2016}. Planck scale is where quantum effects unavoidably enter into the description of gravitational theories. Many quantum gravity theories and phenomenological approaches contemplate quantization of the space-time at that scale. Non-commutative geometrical variables are associated with uncertainty in measuring relative positions or rotation \cite{Hogan17,Hogan18}. It has been argued that light, meanwhile propagating in the arms of an interferometer, would sum up incremental displacement at each discrete Planck interval step leading to a measurable effect. Long-range (spacelike) quantum-like correlation among these variables, and the consequent reduction of the effective numbers of independent degrees of freedom, would allow also to account for the holographic principle. It is indeed in the comparison of the noise among different interferometers that one can find the signature of this ‘exotic’ correlation and provide experimental clues to the nature of those degrees of freedom. The range of this correlation is expected to be only bounded by causality, namely the two interferometers should occupy the same space time volume.
However, since Planck scale effects could have different behavior depending on the geometrical configuration of the arms (radial, rotational, translational symmetry),
a table top or single room experiment is desirable because of the necessity of frequent and quick reconfiguration of the geometry of the system (change of the symmetry/distance). Lack of sensitivity due to the smaller scale (namely one order of magnitude with respect to Fermilab holometer) must be compensated. Increasing the power is possible \cite{note1} 
but only with quantum enhanced techniques the sensitivity of a large experiment can be reproduced or even improved.
If confirmed, holographic noise would provide the first empirical support to theories attempting to unify quantum mechanics and gravitation. At the moment the holometer is operated with classical light only.


According to~\cite{Ruo-Berchera2013, Ruo-Berchera2015} injecting correlated quantum modes such as twin beam (two mode squeezed vacuum), instead of independent beams, allows a drastic improvement of the sensitivity with respect to the classical case, but the advantage can be in principle disruptive also with respect to the use of two independent squeezed modes. The last condition is obtained when the efficiency and the control of the system are such that photon number entanglement of twin beam is efficiently preserved up to the joint detection \cite{Ruo-Berchera2015}.

In the present paper we realize the first proof of principle of this scheme, showing experimentally the advantage of quantum correlation with respect to the classical case \textcolor{black}{in detecting faint signals that can be either correlated or uncorrelated in two interferometers,} and matching the same performance of the double squeezing configuration \textcolor{black}{for the former case}.  For that purpose we use non-classical correlated states in quadratures (twin beam -like state) that provide the same  performance of the entanglement in the actual efficiency regime, reducing at the same time the complexity of the experiment. Our results, on the one hand, open a new branch in quantum metrology, i.e. distributed quantum correlation among several interferometers with great potentiality (provided a large efficiency), on the other hand pave the way for unprecedented sensitivity in devices dedicated to search of Planck scale effects.

\section*{\textcolor{black}{Results}}
\subsubsection*{The experimental setup}

\begin{figure}[th]
	\centering
	\includegraphics[trim={4.5cm 0 0.3cm 0}, clip, width=\columnwidth]{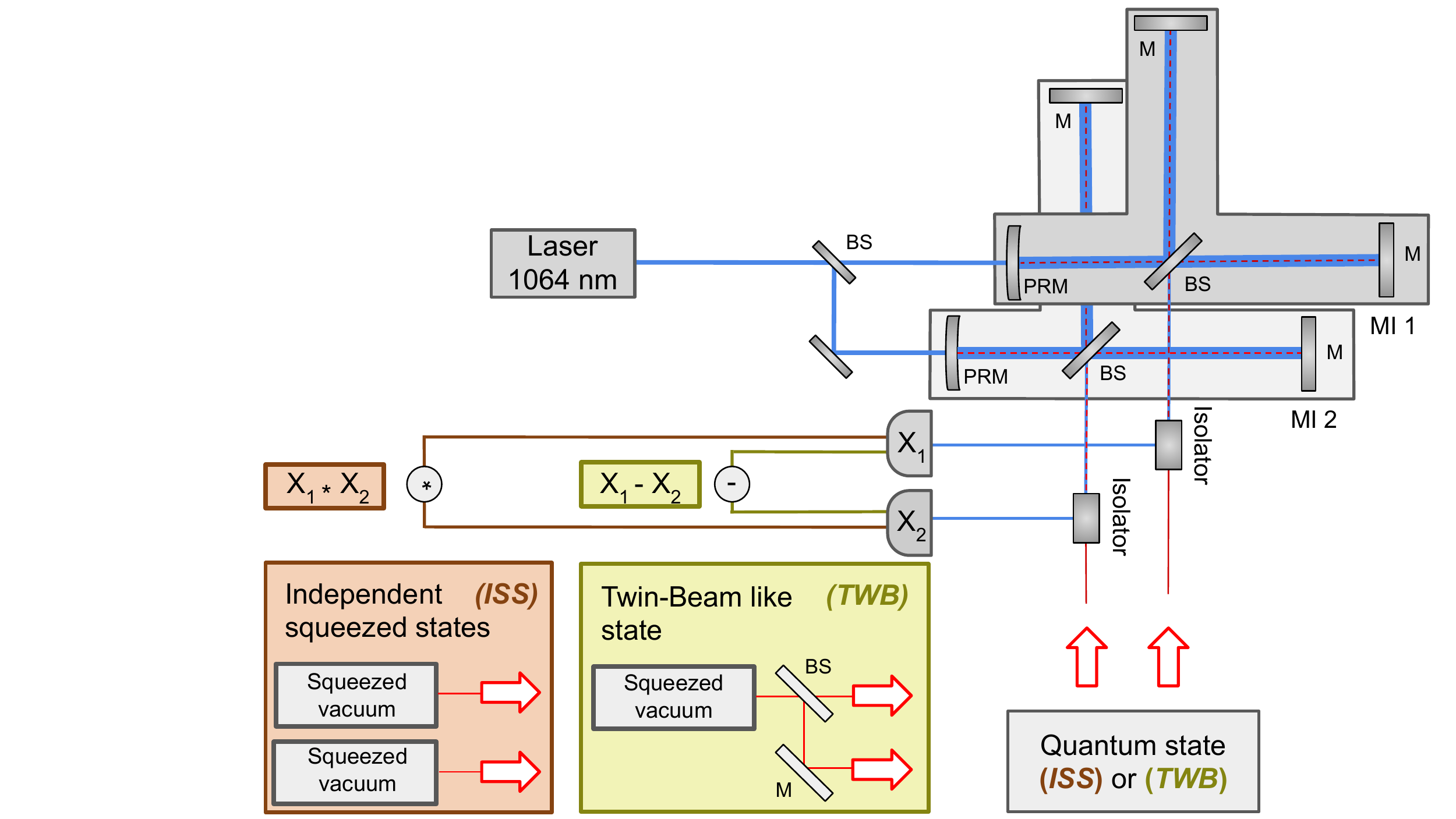}
	\caption{
		Simplified schematic of the double-interferometer setup. Two Michelson Interferometers (MI1, MI2) with arm length $L = \SI{0.92}{m}$ were co-located, with a distance between the two balanced beam splitters (BSs) of around 10 cm.
		M: piezo-actuated high-reflectivity (99.9$\%$) end mirrors.
		PRM: partially reflecting (90$\%$) power recycling mirror, radius of curvature $r_c = \SI{1.5}{m}$.
		$X_1 (X_2)$: read-out signals.
		A Faraday isolator in each output port allowed for measuring the read-out signals while either independent squeezed states \emph{ISS} or twin beam-like states \emph{TWB} were injected into the antisymmetric ports.
		For the \emph{ISS} case the cross-correlation of the outputs is calculated, while for the \emph{TWB} case the output subtraction is the relevant quantity.}
	\label{fig:scheme}
\end{figure}

A simplified schematic of the experiment is shown in Fig.~\ref{fig:scheme}. A similar configuration is also used in large scale experiments, such as LIGO~\cite{Aasi2013} and the Fermilab holometer~\cite{Chou2016}, however, the latter without the injection of quantum states.
Each interferometer consisted of two piezo actuated end mirrors, a balanced beam splitter, and a power recycling mirror. The relative phase of the individual interferometer arms was chosen such that each read-out port was close to the dark fringe and most of the power in the interferometer was recycled~\cite{Meystre1983}.
The interferometer fringe position was chosen such that the output power was $\SI{500}{\micro W}$. The $\sqrt{P}$ scaling of the shot-noise power spectral density at $\SI{13.5}{MHz}$ was verified by increasing the input power in the interferometer (see \textcolor{black}{Methods, locking scheme section, for further details}).

Each MI was fed with 1.5\,mW of 1064\,nm light from a low noise Nd:YAG laser source. The same laser source was used for squeezing generation. 

The squeezed-light sources were based on parametric down-conversion in a potassium titanyl phosphate crystal placed in semi-monolithic linear cavities (see \textcolor{black}{Methods}). The squeezed light was injected in the two MIs via their antisymmetric (read-out) ports. The read-out signals were separated from the squeezed modes by means of optical isolators. The amount of squeezing before injection into the interferometers was measured on a homodyne detector for both sources to be -6.5\,dB relative to the shot noise level (SNL).

The coupled interferometers performance is severely affected by optical losses, which have been carefully characterized (See \textcolor{black}{Methods, optical losses estimation section}).

\subsubsection*{The two possible configurations}
The experiment was performed in two different configurations, as shown in Fig.~\ref{fig:scheme}.
	
	\emph{Independent squeezed states} - \emph{(ISS)}. Two independent squeezed states were injected into the interferometers' anti-symmetric ports. This case represents the benchmark to compare with the twin beam approach.
	A faint correlated \textcolor{black}{stochastic signal} acting in both the MIs can emerge by calculating the cross-correlation of their outputs in the time domain, or the cross-spectrum in the frequency domain, even if in a single interferometer the signal is completely hidden by the shot-noise.
	The quantum noise \textcolor{black}{(shot noise)} in an interferometer is given by the contribution of the vacuum fluctuation entering the unused port of the \textcolor{black}{beam splitter}. If vacuum is substituted by a squeezed state, with a reduction of the noise in one quadrature (at the expense of increasing noise in the other one), and the phase difference with the main laser set properly, the output intensity fluctuation reflects the one of the squeezed quadrature \cite{Schnabel16}. This improves the signal to noise ratio of the phase signal measurement. Thus, the use of two independent squeezed sources, providing a reduction of the photon noise in each interferometer output, while leaving them uncorrelated, also leads to a significant enhancement in the cross-correlation measurement~\cite{Ruo-Berchera2013, Ruo-Berchera2015}.
	
	\emph{Twin beam-like state} - \emph{(TWB)}. A single squeezed state was split on a balanced beamsplitter and the modes were injected into the antisymmetric MI ports.
	The two split beams show non-classical  correlations along one quadrature direction similar to the one present in a proper twin beam or two-mode squeezed state \cite{Schnabel16}. In practice, it turns out in a stronger than classical correlation between the fluctuation of the interferometers output intensities. Exploiting this quantum correlation, we \textcolor{black}{demonstrate} a noise reduction below the SNL in the subtracted outputs of the two interferometers. The subtraction is sensitive to signals that are not correlated between the interferometers, and the obtained noise reduction allows detection of signals of smaller amplitude.
	
	We emphasize that genuine twin beam photon-number entanglement can, in principle, offer far superior performance~\cite{Ruo-Berchera2013, Ruo-Berchera2015}, but, though robust to decoherence~\cite{Benatti2017}, requires efficiency and stability control yet out of reach in realistic systems.
	


\subsubsection{Independent squeezed states (ISS)}

The same stochastic signal \textcolor{black}{(simulating the presence of the holographic noise)} was injected by two electro-optical modulators (EOMs) in both interferometers, with an amplitude well below the sensitivity of the single MI, approximately 1/5 of the SNL.
The cross-correlation is conveniently evaluated in time domain by considering the normalized covariance, defined as:
\begin{equation}\label{rho}
\rho(\tau)=\frac{|\mathrm{Cov}( X_1(t) X_2(t+\tau))|}{\sqrt{\mathrm{Var}( X^{\mathrm{SNL}}_{1}(t)) \mathrm{Var}( X^{\mathrm{SNL}}_{2}(t))}} \ ,
\end{equation}
where $X_1(t)$ ($X_2(t)$) is the time series of the read-out signal of the first (second) interferometer, while $X^{\mathrm{SNL}}_{1}$ ($X^{\mathrm{SNL}}_{2}$) refers to the shot noise limited, classical, signal.

\begin{figure}[th]
	\centering
		\includegraphics[trim={4cm 0cm 13cm 0}, clip, width=\columnwidth]{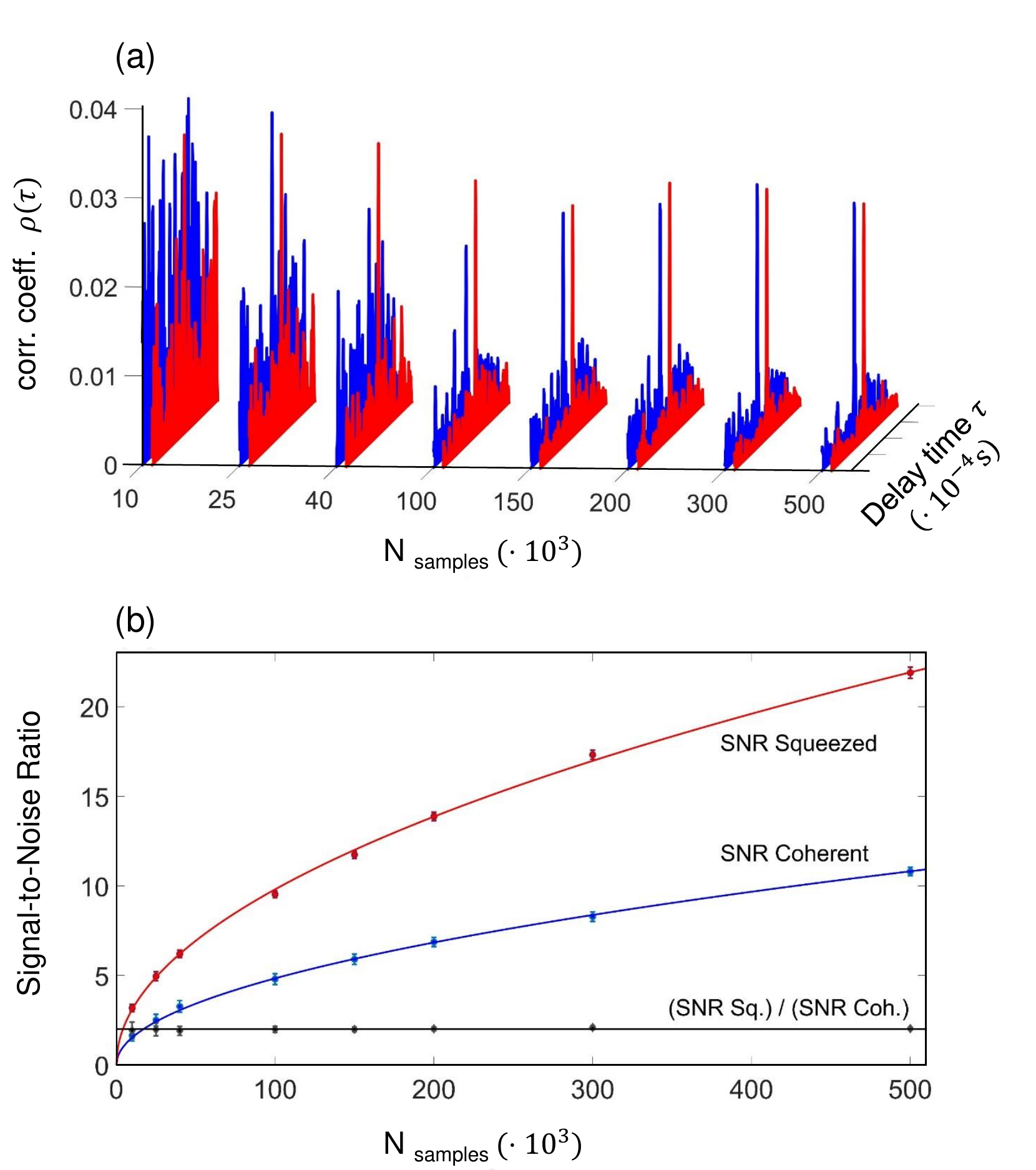}
	\caption{
		Temporal cross-correlation analysis. The red curves correspond to the squeezing injection configuration \emph{ISS} and the blue curves to the coherent case. a) Normalized covariance $\rho(\tau)$ of the data versus the number of samples and as a function of the delay time $\tau$.
		b) Signal to noise ratio (SNR) as a function of the number of samples. Dots represent experimental data (error bars are too small to be appreciated). Data is well fitted by a function proportional to $\sqrt{N_\text{samples}}$ (blue and red  traces). Black data is the ratio between squeezed and coherent SNR values.
	}
	\label{fig:ind_sq_time}
\end{figure}

Figure~\ref{fig:ind_sq_time}(a) shows $\rho(\tau)$ of the read-out signals as a function of the number of samples.
The sampling rate was 500 kS s$^-1$ and the total acquisition time was 1 s. Fig.~\ref{fig:ind_sq_time}(a) is plotted by calculating $\rho$ for increasing subsets of the total number of samples. One can observe that the background level systematically decreases with increasing number of samples, as expected for statistically independent noise sources, while the peak height due to the injected correlated signal approaches a constant value (apart some random fluctuations).
The correlated noise peak, which is initially hidden in background noise, is resolved for shorter integration times when squeezing is used (red traces), compared to the classical case of no squeezing (blue traces). This results in a more rapidly increasing SNR in the squeezing case, as shown in Fig.~\ref{fig:ind_sq_time}(b).
Here, each data point is derived from the ratio between the normalized covariance peak height for $\tau=0$ and the floor level, averaging over 19 independent data sets similar to the one originating Fig.~\ref{fig:ind_sq_time}(a). The SNR with squeezing injected is consistently higher, by a factor of 2. \textcolor{black}{This is in accordance with the squeezing level introduced and the level of losses estimated inside the interferometer as discussed in Methods, optical losses estimation section.}
Note that the SNR scales as the square root of the number of samples, i.e. with the square root of the acquisition time. Thus, a factor 2 of SNR enhancement corresponds to a reduction of 4 times in the measurement time.

\begin{figure}[th]
	\centering
		\includegraphics[trim={4.5cm 9cm 4cm 3cm}, clip, width=\columnwidth]{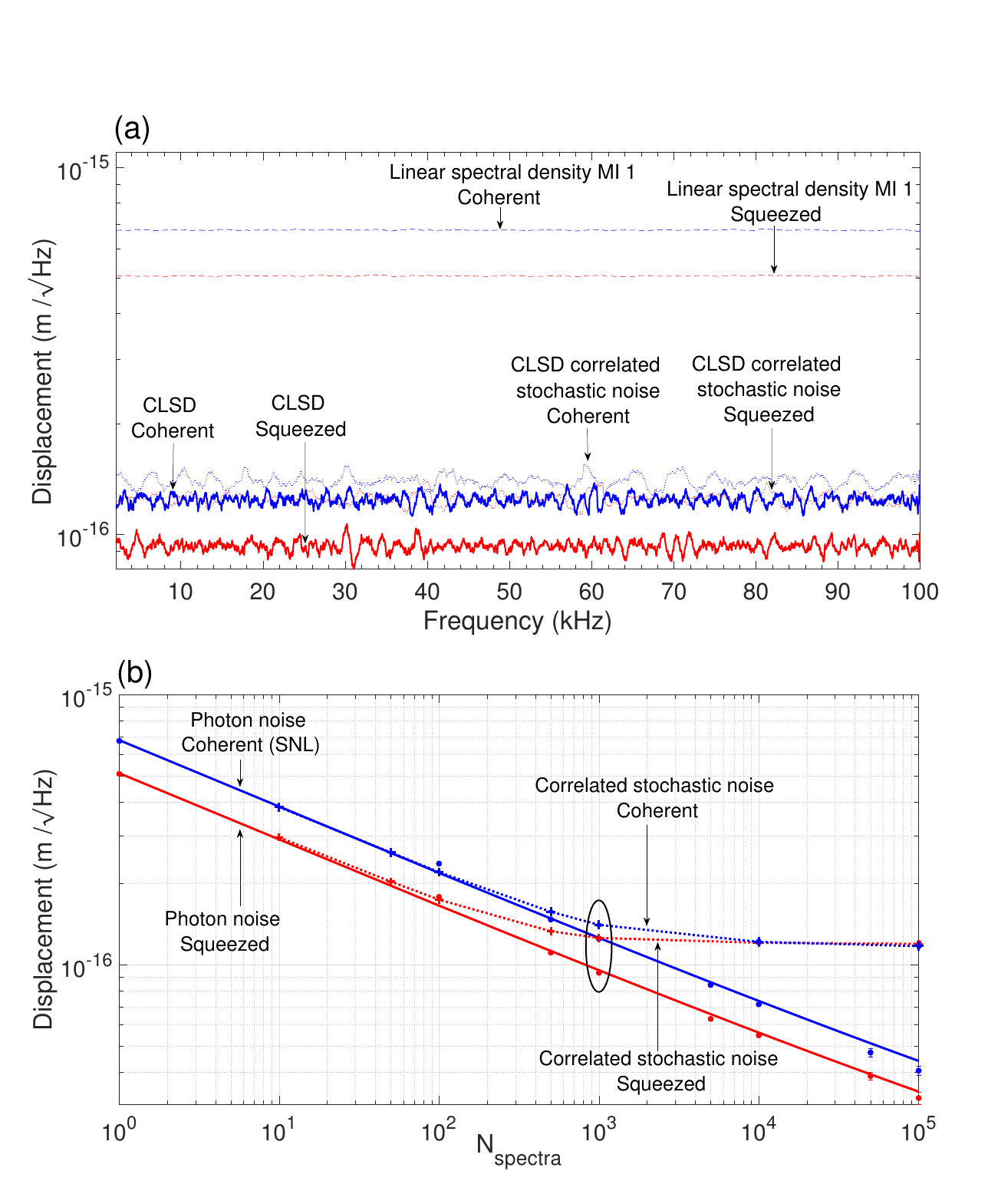}
	\caption{
		Cross-Linear Spectral Density (CLSD). Red curves refer to the quantum-enhanced case \emph{ISS} and blue curves to the classical coherent one.  a) The CLSD  of the photon noise floors, for $N_\text{spectra}=1000$, are plotted for the squeezed (thick  red line) and coherent (thick  blue line) cases. Faint lines refer to the addition of a stochastic phase signal of amplitude 1/5 of the shot-noise-limit (SNL). The linear spectral densities of the individual interferometer (with and without squeezing injection) are plotted for reference. The spectra were calculated from the read-out signals down-mixed at 13.5\,MHz. b) The average level of the CLSD is shown as a function of $N_\text{spectra}$. The data inside the black circle refers to the average of the CLSD in (a). Thick solid lines are the fitting curves confirming the scaling with $N_\text{spectra}^{-1/4}$. Error bars are also reported (for most of the points they are too small to be appreciated).
	}
	\label{fig:CPSDsqxsq}
\end{figure}

In the spectral domain, correlated signals can be extracted by the cross-linear spectral density (CLSD) of the two interferometers, as shown in Fig.~\ref{fig:CPSDsqxsq}. This quantity is obtained by dividing the time series in $N_\text{spectra}$ bins. For each bin the cross-power spectral density is calculated as the discrete Fourier transform~\cite{Oppenheim1999} of the cross-correlation. The average of the $N_\text{spectra}$ cross-power spectral density values is then evaluated. Note that in analogy with the cross-correlation, the average reduces the contribution of the uncorrelated \textcolor{black}{photon noise} by a factor $N_\text{spectra}^{-1/2}$, while the correlated \textcolor{black}{signal} is unaffected. To obtain the CLSD the square root of the power spectral density is calculated: therefore the overall scaling of the uncorrelated contribution with the number of spectra is $N_\text{spectra}^{-1/4}$. For $N_\text{spectra}$ sufficiently high, the CLSD approaches the linear spectral density of the correlated signals.

Figure~\ref{fig:CPSDsqxsq}(a) shows the CLSD in a bandwidth of 100\,kHz after down-mixing \cite{note2} 
the detected signal at 13.5\,MHz. The acquisition time was 20\,s ($10^6$ samples) and the average was performed over $N_\text{spectra} = 1000$. The CLSDs of the solely uncorrelated photon noise are reported as thick solid lines, red with squeezing injection and blue without. They represent the sensitivity level in the detection of \textcolor{black}{a} correlated \textcolor{black}{signal}, calibrated in \textcolor{black}{m Hz}$^{-1/2}$ (see \textcolor{black}{Methods, sensitivity calibration section}). This should be compared with the respective sensitivity of the single interferometer in the two cases (dashed lines). While almost a factor of 5.6 of improvement is gained by the cross-spectra statistical averaging, an additional multiplicative factor of 1.35 (corresponding to about 2.6 dB) is obtained from the injection of squeezed states.

Injecting a small stochastic signal, the corresponding CLSDs (faint CLSD traces in Fig.~\ref{fig:CPSDsqxsq}(a)) are almost overlapping the CLSD of the photon noise in the coherent (classical) case, while the traces are clearly separated from the CLSD of photon noise when squeezing is applied.

Figure~\ref{fig:CPSDsqxsq}(b) shows the scaling of the CLSD average with the number of spectra. The CLSDs of the photon noise, independent in the two interferometers, scale as $N_\text{spectra}^{-1/4}$ as expected. \textcolor{black}{When a correlated stochastic signal is injected (simulating the presence of holographic noise), the CLSD reaches a plateau determined by the amplitude of the signal. The maximum absolute sensitivity, i.e. the residual uncorrelated photon noise level, achieved with independent squeezed states is measured} to be $3 \times 10^{-17}$ \textcolor{black}{m Hz}$^{-1/2}$ (with 68\% confidence interval), corresponding to $1/20$ of the SNL. This number also represents a limit to the magnitude of \textcolor{black}{a} correlated \textcolor{black}{signal (e.g. holographic noise) detectable} in this frequency band. \textcolor{black}{We note that, in principle, it is possible to beat the Fermilab holometer performance with a 1 meter long arm, for example considering a higher yet realistic optical power (namely 10 kW), longer averaging time (larger than 1 h) and with the help of squeezing.}

\subsubsection{Twin beam-like state (TWB)}

With the two modes of an evenly split squeezed beam injected into the MIs antisymmetric ports, the non-classical quadrature correlation is expected to provide a \textcolor{black}{photon} noise reduction in the read-out signal subtraction.
Any phase difference between the two MIs produces a change in the relative photo-currents which is detected with sub shot-noise sensitivity.

\begin{figure}[ht]
	\centering
	\includegraphics[trim={2.6cm 0 4cm 0}, clip, width=\columnwidth]{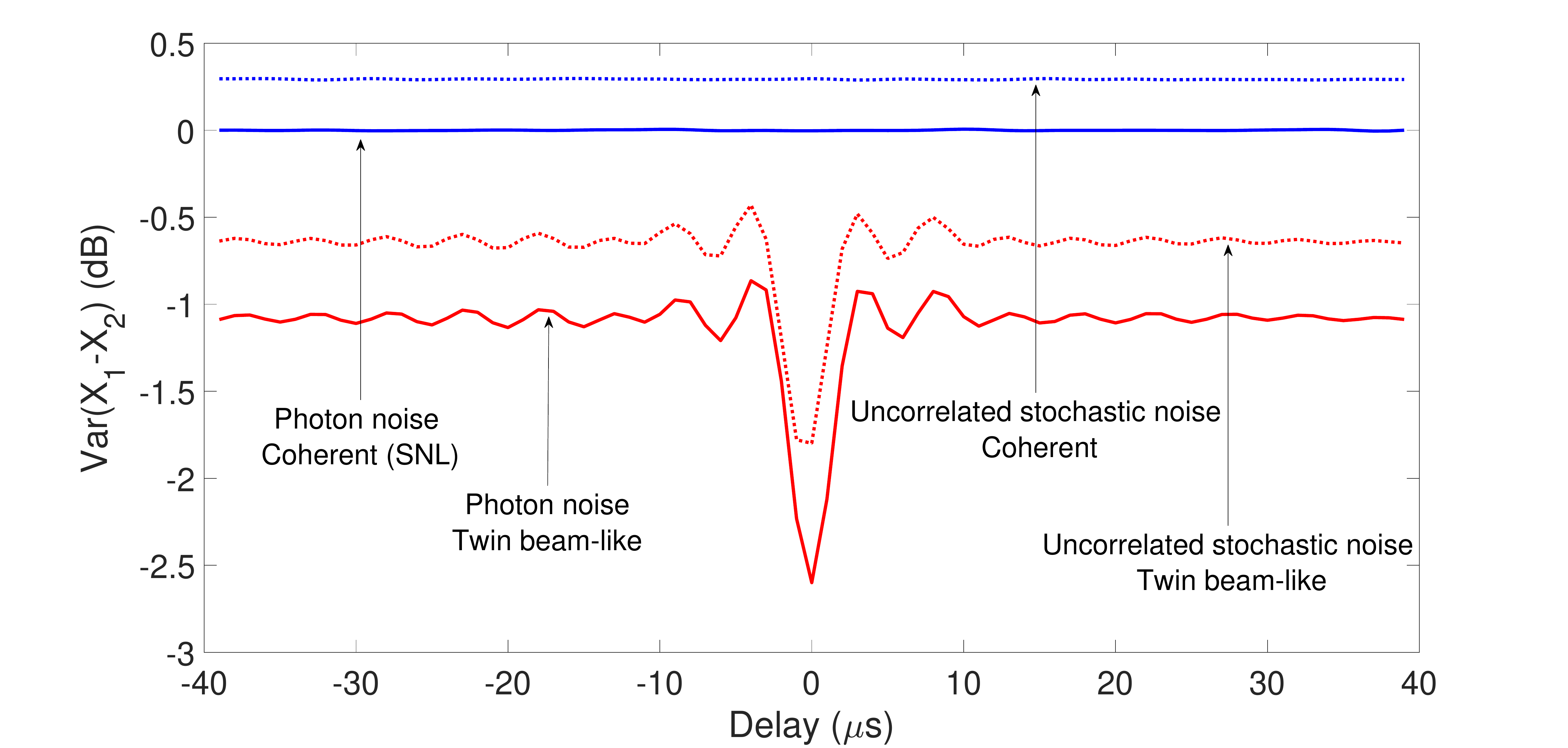}
	\caption{
		Variance on the read-out signals subtraction with varying relative time delay. Red curves refer to the twin beam - like case \emph{TWB} and blue curves to the classical coherent one. Faint lines refer to the addition of an uncorrelated stochastic noise in the two interferometers.
	}
	\label{fig:varX-Y}
\end{figure}

Figure~\ref{fig:varX-Y} shows the variance of $X_1(t)-X_2(t+\tau)$ as a function of time delay $\tau$. More specifically, $\mathrm{Var}(X_1(t) - X_2(t+\tau))= \mathrm{Var}(X_1) + \mathrm{Var}(X_2) -2 \mathrm{Cov}(X_1(t) X_2(t+\tau))$. For $\tau = 0$ the correlation between the two modes leads to a \textcolor{black}{photon} noise reduction of 2.5 dB with respect to the SNL, represented by a dip in the variance (thick  red line). When two uncorrelated  stochastic signals are injected in the MIs, the dip reduces by $\sim$ 1 dB, as shown by the red faint line.
This must be compared with the change between classical coherent trace levels with and without signal injection (blue thick and faint line respectively), which is only $\sim$ 0.3\,dB.
Note that for $\tau \neq 0$, the covariance of $X_1(t)$ and $X_2(t+\tau)$ is zero.
Since $\text{Var}(X_1(t) - X_2(t+\tau))$ preserves some sub-shot-noise features (around 1\,dB), each of the two modes is a squeezed beam affected by 50\,\% loss induced by the beam splitter: the individual squeezing level is therefore degraded, but still present in the global state.

\begin{figure}[ht]
	\centering
	\includegraphics[trim={2cm 0 3.2cm 0cm}, clip, width=\columnwidth]{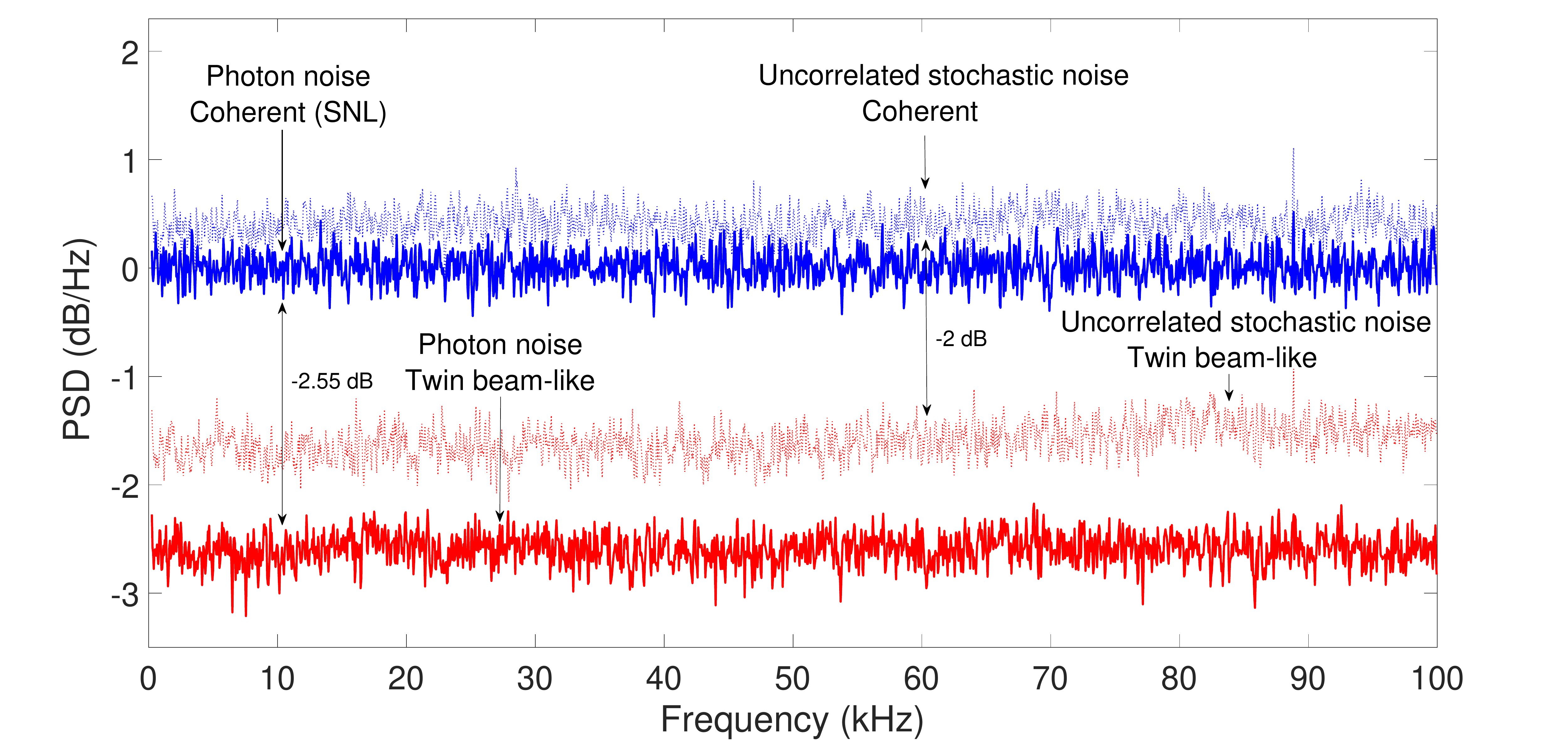}
	\caption{
		Power Spectral Density (PSD) of the read-out signals subtraction calculated from the read-out signals down-mixed at 13.5\,MHz. Red curves refer to the twin beam - like case \emph{TWB} and blue curves to the classical coherent one. Faint lines refer to the addition of an uncorrelated stochastic noise in the two interferometers.
	}
	\label{fig:CPSDtwb}
\end{figure}

This enhancement is also observed in the power spectral density of the subtracted interferometer outputs, plotted in Fig.~\ref{fig:CPSDtwb}. This experimentally demonstrates that the presence of faint uncorrelated \textcolor{black}{signal} can be more easily detected by twin beam-like correlations. Conversely, \textcolor{black}{a} correlated \textcolor{black}{signal} is completely suppressed by the subtraction.
Therefore, measuring the read-out signals subtraction for varying distances between the MIs, or the temporal delay, can provide an alternative way to study the coherence properties of the noise sources under investigation.

\begin{figure}[ht]
	\centering
		\includegraphics[trim={4cm 0cm 4cm 0cm}, clip, width=\columnwidth]{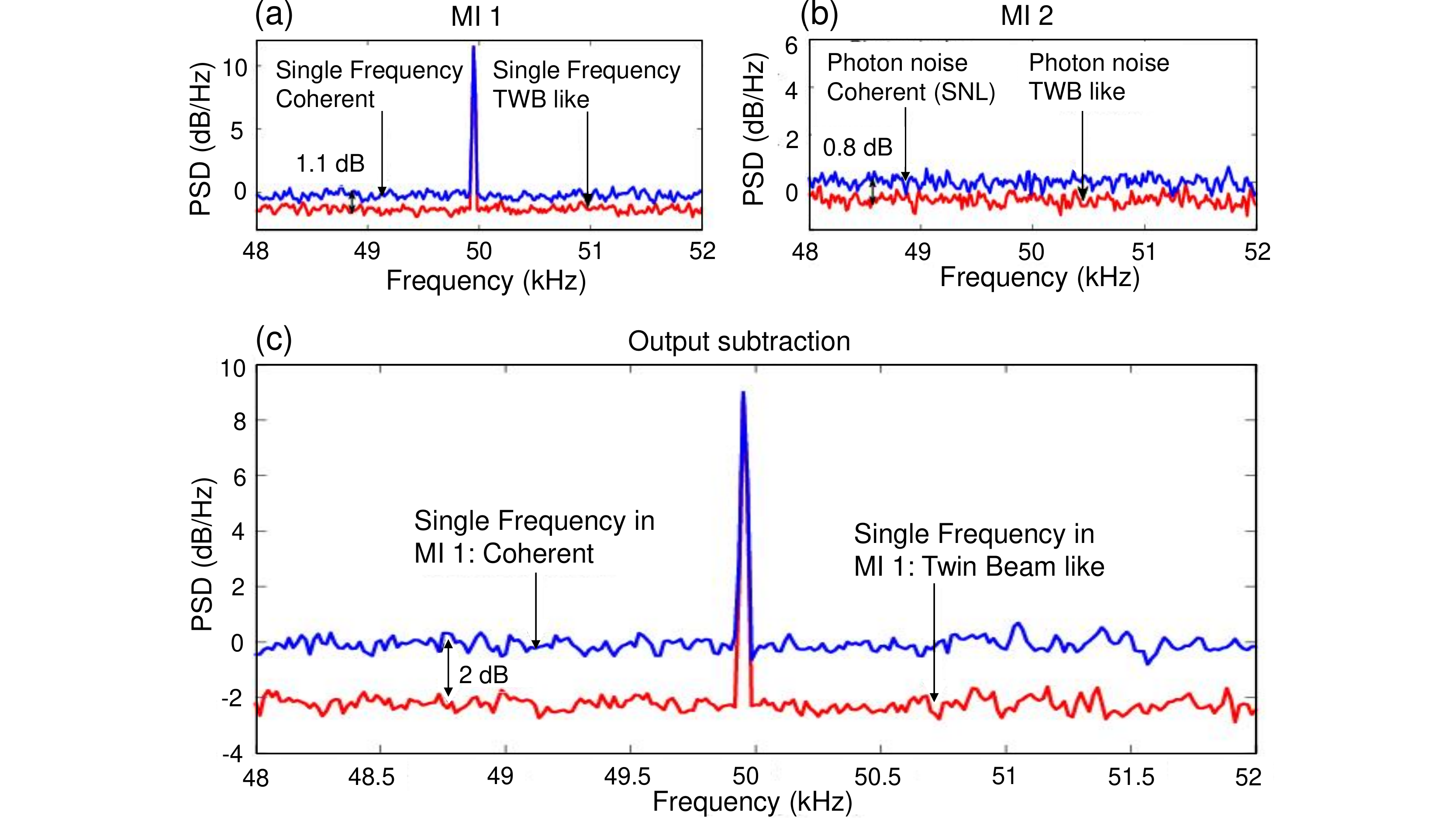}
	\caption{
		Single-frequency tone injected in the first interferometer at 13.55\,MHz. (a),(b): Power spectral densities (PSDs) of the read-out signal in each Michelson interferometer (referred as MI 1 and MI 2) for the coherent case (blue curve) and  when the twin-beam like state (TWB like) is injected (red curve). A 1.1 dB and 0.8 dB quantum-enhancement for MI 1 and MI 2 respectively is demonstrated. (c): PSD of the read-out signals subtraction. The correlation between the two modes leads to 2 dB of squeezing. The frequency axis of all plots correspond to the recorded data after down-mixing at 13.5\,MHz.
	}
	\label{fig:SF_PSD}
\end{figure}

Figure~\ref{fig:SF_PSD} shows the power spectral density of  $X_1(t) - X_2(t)$ for a single frequency tone applied to one of the MIs. Also in this case the quantum-enhancement is clearly visible. The power spectral density enhancements of the individual interferometers are 1.1 and 0.8\,dB respectively, resulting in a collective enhancement of 2\,dB in the output subtraction, facilitated by the non-classical correlation among the modes. This enhancement might be applied to identify \textcolor{black}{unwanted} uncorrelated \textcolor{black}{background} sources, such as scattering or  resonances~\cite{Steinlechner2013}.

\begin{figure}[ht]
	\centering
	\includegraphics[trim={3cm 0 4cm 0}, clip, width=\columnwidth]{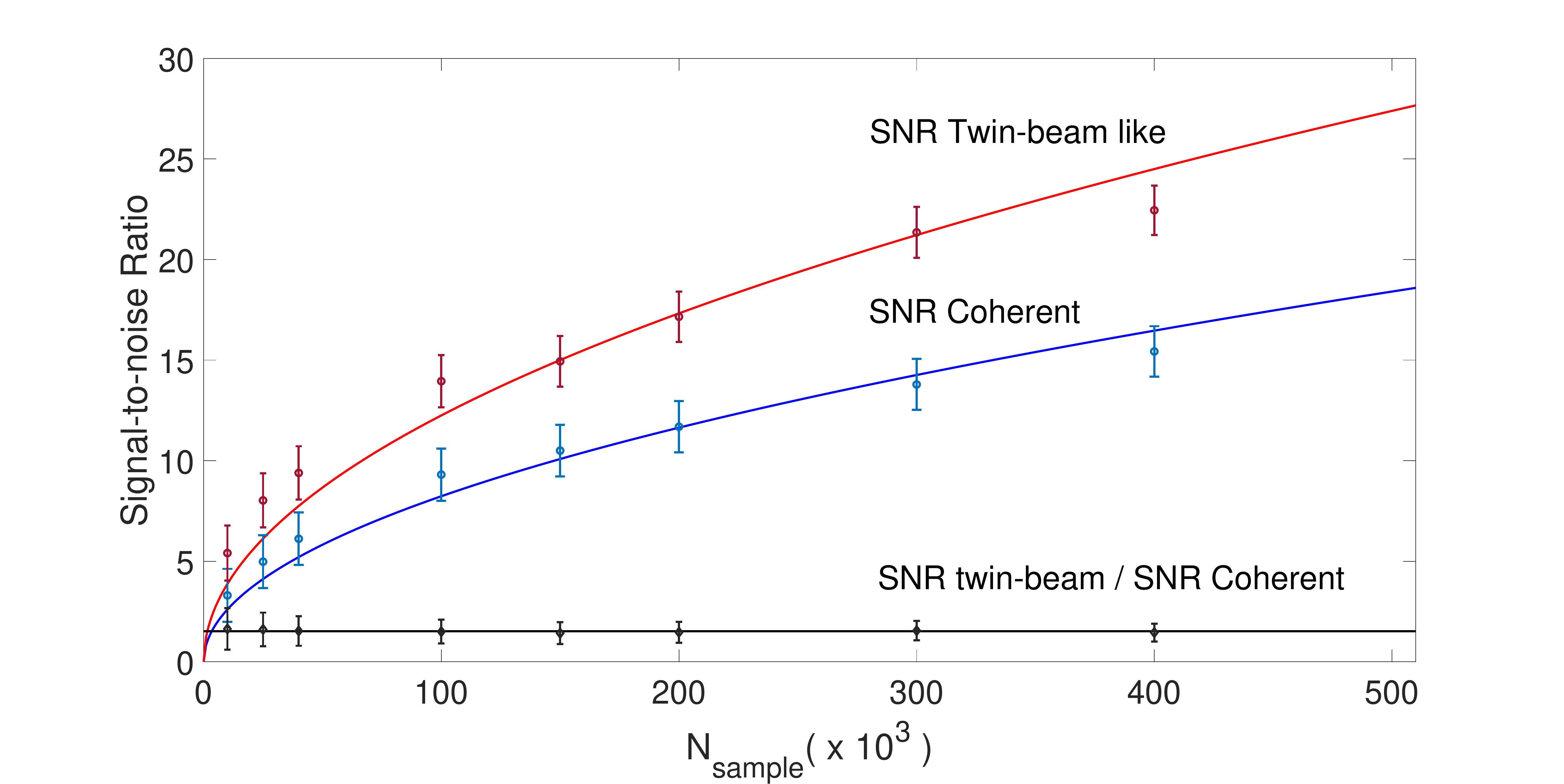}
	\caption{Signal to noise ratio (SNR) of the covariance measurement with twin-beam-like state as function of the number of samples. The red data correspond to the twin beam injection \emph{TWB} and the blue data to the coherent case. Uncertainty bars assume normal distribution of the data and the fit function is proportional to $\sqrt{N_\text{samples}}$. Black data is the ratio between squeezed and coherent SNR values.}
	\label{SNR_twb}
\end{figure}

The twin beam approach also allows an enhanced estimation of \textcolor{black}{a} correlated \textcolor{black}{signal} if one considers the measurement proposed in \cite{Ruo-Berchera2013, Ruo-Berchera2015}, requiring the comparison of two configurations of the setup: one  with correlated \textcolor{black}{stochastic signal} and the other one with uncorrelated \textcolor{black}{signal}. The latter one represents the real background reference, as the \textcolor{black}{holographic} noise of fundamental origin cannot be turned off, and also allows discriminating th\textcolor{black}{is} fundamental \textcolor{black}{signal} from other sources of technical noise that are insensitive to the geometrical configuration. \textcolor{black}{In particular}, the holographic noise would become independent in the two interferometers if the geometrical configuration is changed (for example if one of the arms is rotated of 180 degrees)\cite{Hogan2012,Hogan17}. We note that, under the approximation in which the signal $X$ in each device can be written as the sum of  photon noise and the stochastic white noise \textcolor{black}{(in this case representing the signal)}, $X_{i}=X_{\mathrm{WN}i}+X_{\mathrm{PN}i}$ $(i=1,2)$, it turns out that
\begin{eqnarray}\label{Cov_twb}
\mathrm{Var}(X_1(t)-X_2(t))_{\mathrm{corr}}-\mathrm{Var}(X_1(t) - X_2(t))_{\mathrm{uncorr}}\\\nonumber
=2\mathrm{Cov}(X_{\mathrm{WN}1}(t), X_{\mathrm{WN}2}(t)),
\end{eqnarray}
providing a way to estimate the covariance of the stochastic \textcolor{black}{signal}. Since the photon fluctuation of each term in the left hand side of Eq. (\ref{Cov_twb}) is below the shot noise level (when quantum correlations are injected), we expect a reduction in the uncertainty on the covariance evaluated in this way. As demonstrated in \cite{Ruo-Berchera2015}, when the photon number correlation (entanglement) is injected and preserved at the output ports (requiring very high efficiency and to be very close to the dark fringe), the advantage of this scheme can be disruptive\textcolor{black}{. Considering the best squeezing factor obtained up to now, of 15 dB \cite{Vahlbruch2016}, and an optical power of 10 kW as currently used in large scale  experiments,  an overall efficiency larger than 99$\%$ would be required to have a significant advantage with respect to the independent squeezing setting. The fringe stability strongly depends on the optical power injected and other parameters. For example, for the optical power used in our experiment it is necessary to reach $<$ 0.1 $\mu$rad. N}evertheless a significant \textcolor{black}{quantum} advantage persists also for less demanding conditions. \textcolor{black}{For our experimental parameters, theory predicts almost the same quantum enhancement as for the independent squeezing case \cite{Ruo-Berchera2015}.} \textcolor{black}{However we stress that the twin beam} approach is different from using independent squeezing, both conceptually and for potential perspectives. Fig. \ref{SNR_twb}  presents the SNR obtained in our experiment when the same level of white noise is injected, first correlated and then uncorrelated, to perform the variance subtraction in Eq. (\ref{Cov_twb}). Each point, corresponding to a fixed number of samples (measurement time),  is obtained by calculating the ratio of the mean value of the covariance and its standard deviation  in a statistically significant number of subsets. The red series corresponds to the injection of a twin beam-like state while the blue one stands for the coherent case. The quantum advantage, as extracted from the fit, is 1.52 which corresponds to about  1.8 dB, quite close to the 2 dB reported for example in Fig. \ref{fig:CPSDtwb} in presence of uncorrelated noise. The results of Fig. \ref{SNR_twb}, can be in principle compared with the ones reported in Fig.~\ref{fig:ind_sq_time}(b) related to the measurement of the normalized covariance in the independent squeezing case (the normalization is irrelevant for the SNR). However, we stress that some  technical difference occurred in the realization of the two measurements, in particular the starting level of squeezing was of 2.5 for twin beam-like case and 3dB for independent squeezing due to alignment, and also the injected noise was actually higher in the twin beam-like case. It can be shown numerically that these discrepancies account for the different quantum advantage obtained in the two cases. Thus, \textcolor{black}{apart} from technical issues that are always subject to fast technological evolution, the twin beam correlation has been demonstrated for the first time to be useful in an interferometric scheme with potential important applications. In fact, this represents a first step that paves the way to achieve the dramatic improvement predicted in \cite{Ruo-Berchera2013, Ruo-Berchera2015}.

\section*{\textcolor{black}{Discussion}}

In this work we have presented the first realization of a double interferometric system enhanced by quantum correlations, where each interferometer is injected with a mode of a bipartite twin beam-like state. We have demonstrated its applicability to the measurement of possible Planck scale effects~\cite{Chou2017, Hogan2012, Chou2016} expected in the MHz-range. The technique proposed here can be in principle implemented also at audio frequencies, more interesting for stochastic gravitational wave background, although in this case it is necessary taking into account the quantum back-action, that is not \emph{a priori} negligible in that frequency range\textcolor{black}{. Although an extended discussion of this point is largely beyond the purpose of this work we note that a possible approach to back-action noise reduction can be found in  \cite{Ma17,Sudbeck19,JetYap19}.}

The twin beam-like state provides a clear quantum enhancement with respect to both the single interferometer and a system of two of them fed with only coherent beams. We have also compared its performance with two independent squeezed states injected in the same system, that represent the state of the art in the field, substantially obtaining a similar behavior in the detection of a correlated signal. \textcolor{black}{In perspective, considering state of the art squeezing generation (15 dB) and an overall efficiency of 90$\%$, the expected sensitivity will be improved by almost one order of magnitude with respect to a pure classical approach, such as the one used at Fermilab holometer.} Incidentally, in our experiment we put a limit to certain fundamental effects related to Plan\textcolor{black}{c}k scale physics at frequency of 13.5 MHz. However, as predicted in~\cite{Ruo-Berchera2013, Ruo-Berchera2015}, the twin beams approach, representing the first instance of a conceptually new class of quantum enhanced interferometry, opens the perspectives of an extremely high advantage in the case of large detection efficiencies and fine control of the fringe point, still challenging in present technology.

\section*{\textcolor{black}{Methods}}

\subsubsection{Locking scheme}

\begin{figure}[th]
	\centering
	\includegraphics[trim={0.6cm 0 2.7cm 0}, clip, width=\columnwidth]{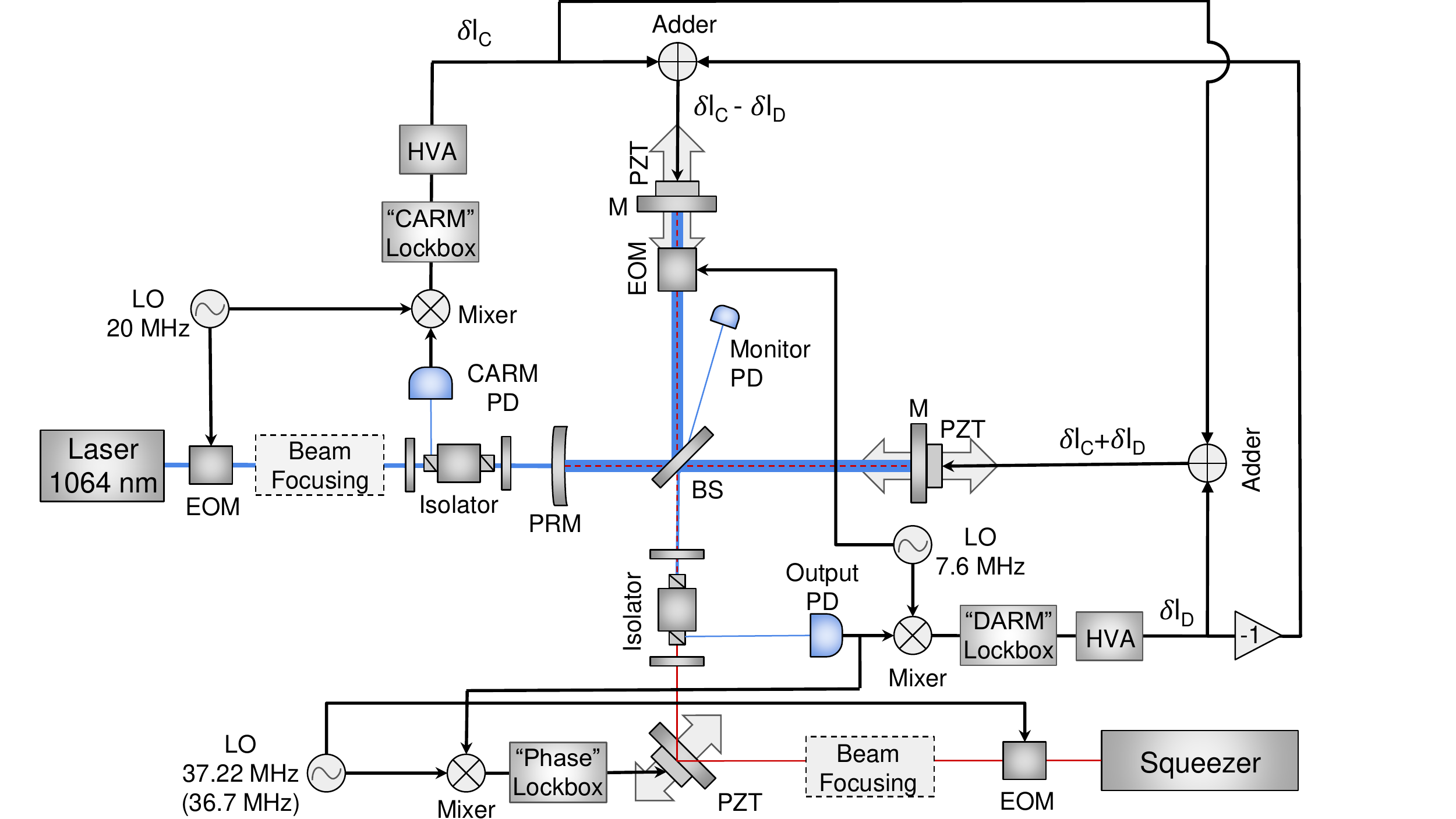}
	\caption{Detailed schematic of the individual power-recycling Michelson interferometer (MI).  The MI is fed with a Nd-YAG laser of wavelength $\lambda= 1064$ nm.  The input power is maintained around $P_{\mathrm{in}}$= 1.5 mW. 
	 Each power-recycling MI is formed by a power-recycling mirror (PRM) of 90$\%$ reflectivity 
	 , a 50-50 beam splitter (BS),  and  two  plane end-mirrors M (reflectivity $\sim$ 99.9$\%$). These mirrors are mounted on piezoelectric actuators (PZT), used for maintaining the interferometer at the desired working point through the implemented locking scheme (see Methods, locking scheme section). An electro-optical modulator (EOM) in one arm of the MI is eventually used to inject a phase signal. Two Faraday isolators permit to separate the light exiting the interferometer from the entering quantum states.
	The  acquisition  board  consists  in  a  14  bit  data  acquisition  system  with  4 channels (DC and AC for each MI) and acquisition rate 500 kSample/s.  The DC signal from the output PD is collected without any further processing, while the AC signal is down mixed at 13.5 MHz, preamplified and low passed at 100 kHz.
	HVA: high-voltage amplifier. LO: local oscillator. PD: photo-diode, InGaAs with high quantum-efficiency photodiodes (99$\%$) and low noise (Noise Equivalent Power $1.2 \times 10^{-11}$ \textcolor{black}{W Hz}$^{-1/2}$) are used at the read-out port. Beam-Focusing: set of lenses in order to have Gaussian Optical mode TEM$_{00}$. $\delta l_C, \delta l_D$: correction signals from the common arm length (CARM) and differential arm length (DARM) lockbox respectively.}
	\label{fig:scheme_single}
\end{figure}

A detailed schematic of the individual interferometer is shown in Fig.~\ref{fig:scheme_single}.

Because of the power recycling mirror (PRM), a change in any individual arm length affects the interferometer power output and the cavity resonance condition at the same time. Therefore two degrees of freedom need to be controlled: rather than considering the individual arm lengths $L_1$ and $L_2$, we define the common arm length (CARM) as $(L_1+L_2)/2$ and the differential arm length (DARM) as $L_1-L_2$. The CARM controls the cavity resonance, while the DARM controls the interferometer fringe position.

The CARM was locked to the cavity resonance through the Pound-Drever-Hall locking technique (PDH) \cite{Drever1983}. A 20 MHz phase modulation was applied to the input beam by an electro optical modulator (EOM).
The beam reflected from the PRM was separated from the incoming beam with a Faraday isolator and detected by a photo detector which generated an error signal by demodulating the signal at 20 MHz. This error signal was processed by a proportional-integral (PI) controller, and the output was applied to both end mirror actuators with no relative phase lag.

The DARM was locked close to the interferometer dark port using a phase modulated sideband at 7.6 MHz. The read-out signal was demodulated at 7.6 MHz and processed by a PI. The output of this PI controller was applied to the end mirrors actuators differentially (i.e.\ the two actuators moved exactly out of phase).

The DARM locking sideband at 7.6 MHz was generated by an EOM, positioned in one arm of the interferometer. This EOM was also used to generate single frequency tone and stochastic noise modulation from 12.3 to 13.8 MHz (extending well outside the measurement band). The stochastic noise generator had two channels and allowed producing both uncorrelated and highly correlated noise between them. These modulations acted as artificial signals for testing the system performance.

The intra-cavity power was measured using a spurious reflection from the beamsplitter. This was used to estimate the gain of the cavity, which in typical working conditions varied between 8 and 10.

\subsubsection{Sensitivity calibration}
The sensitivities plotted in Fig.~\ref{fig:CPSDsqxsq} are expressed in \textcolor{black}{m Hz}$^{-1/2}$. This was obtained by calibrating the strain $\delta x$ produced by the phase modulator generating the phase noise:

\begin{equation}
\delta x (\mathrm{m} \cdot \mathrm{Hz}^{-1/2})=\frac{\lambda}{2}\frac{V_\mathrm{rms}}{V_\pi}\frac{1}{\sqrt{BW}} \ ,
\end{equation}

where $\lambda = \SI{1064}{nm}$, $BW$ is the measurement bandwidth, and $V_\pi$ and $V_\mathrm{rms}$ are, respectively, the half-wave voltage and the input root-mean-square voltage at the phase modulator.

\subsubsection{Squeezing generation}

\begin{figure}[ht]
	\centering
	\includegraphics[trim={2cm 0cm 2cm 0cm}, clip,  width=\columnwidth]{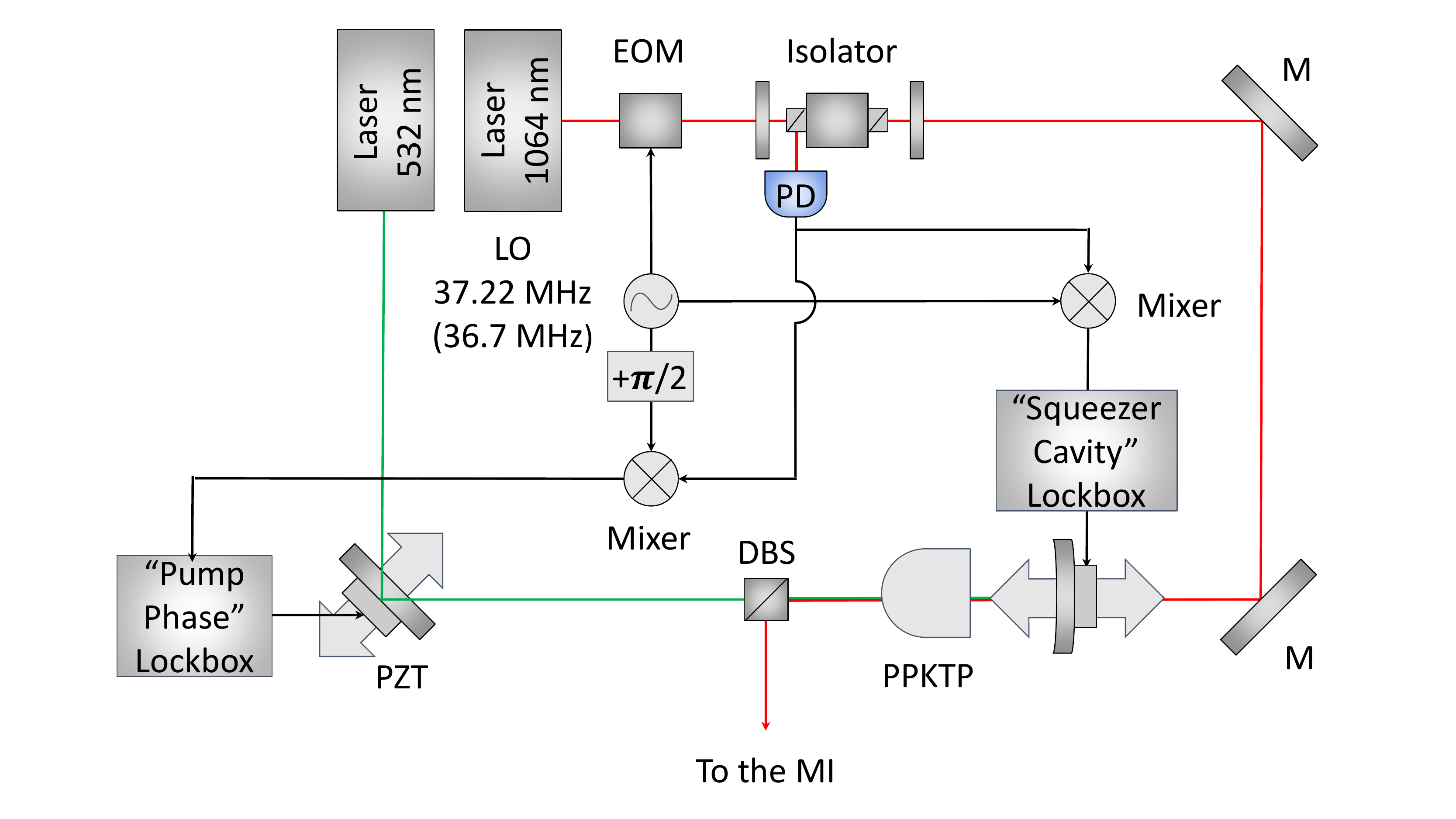}
	\caption{Schematic of the squeezed light source. PPKTP: potassium titanyl phosphate crystal. DBS: dichroic beam splitter. PZT: piezoelectric actuators. EOM: electro-optical modulator. LO: local oscillator. PD: photo-diode. M:mirror. MI: Michelson interferometer}
	\label{fig:squeez}
\end{figure}

The schematic of the squeezed-light source is shown in Fig.~\ref{fig:squeez}. The source was based on parametric down-conversion in a potassium titanyl phosphate crystal placed in a semi-monolithic linear cavity.

The cavity was seeded with 1064 nm light to lock the cavity length, using a PDH lock, and the crystal was pumped with 70 mW of 532 nm light. The pumping gave rise to a phase dependent amplification of the seed beam. Locking this relative phase between seed and pump to the bottom of the generated gain curve, through a sideband generated error signal, the squeezing was produced in the amplitude quadrature.

\subsubsection{Quantum Noise Locking}

The outputs from the squeezing sources were injected from the antisymmetric port into each interferometer, according to configuration \emph{ISS} or \emph{TWB} in Fig 1. The control beams of both squeezing sources entered the interferometers with the squeezed light carrying sidebands at 37.22 MHz (36.7 MHz), see Fig.~\ref{fig:scheme_single}. These sidebands were used to lock the squeezing phase to the bright output beam of the interferometer through a piezo actuated mirror in the squeezed beam path.

\subsubsection{Data Acquisition}

The alternating current (AC) outputs of the photo detectors were demodulated at 13.5 MHz and lowpass filtered at 100 kHz. Both the two direct current (DC) output signals and the two demodulated AC output signals (dubbed $X_1$ and $X_2$ in the text) were simultaneously recorded using a data acquisition card. When correlated white noise or twin beam-like light was injected, the phase between the two electrical local oscillators used for demodulation was adjusted to maximize the cross correlations.

\subsubsection{Optical losses estimation}

We have performed a characterization of the losses of the main coherent beam based on a model of the fringe of a power recycling interferometer. By fitting the experimental data of the gain $G$ and the power measured at the anti-symmetric port $P_{AS}$ as functions of the DARM with the theoretical expression, one can extract both the reflection of the PRM and the mode mismatch:

	Reflection of the PRM: $R_{prm}=0.91(1)$. This value, obtained from the fit, is consistent with an independent measurement and with the nominal value provided by the producer, supporting the validity of the model;\\
	
	Mode mismatch and other losses inside the power-recycling cavity (excluding $R_{prm}$), and from the beam splitter to the detector included: $L_d=0.26(1)$. This accounts for cavity mode mismatching, the presence of the EOM inside the arms and output isolators (2 of them are used in a single interferometer).\\

The other optical losses from end mirror and the beam splitter are negligible.

From the value reported above we have an overall efficiency of $0.91*(1-0.26) =0.67$, i.e.  $33\%$ of losses. On the other hand, the observed reduction of squeezing from 6.5 dB to 2.6 dB corresponds to a loss of about $42\%$. The residual $9\%$ of difference, can be attributed to the double passing through the output isolators (injection and readout) of the squeezed beam, while the local oscillator passes only once. \textcolor{black}{It is worth observing that the reported value of losses can slightly change from one run to another because of different misalignment and accordingly also the level of quantum noise reduction in the various measurements.}

\section*{Data Availability}

All relevant data that have been acquired to generate the figures presented in here and to obtain the parameters reported are available at least from the authors.


\begin{thebibliography}{10}
\bibitem{note1}{Of course, in principle it is always possible to reduce the shot noise of a factor $S$ by increasing the optical power of a factor $S^2$ (which is not a favorable scaling). In practice, other sources of technical noise, detrimental for the sensitivity, become more and more important with the increasing the beams energy. These include thermal effects on the optics material (beam splitter, mirrors etc…), spurious scattering from optics imperfections and residual gas impurities, even from photons on the tails of the Gaussian beam again the side walls.}
\bibitem{note2}{Down mixing at a certain frequency $f$ refers to the procedure of beating of the electric signal with a local oscillator at frequency $f$ by a mixer. It allows a conversion towards lower frequencies of the signal, in our case for the purpose of a temporal sampling, which would be challenging at the original frequency range due to limitation of digitization electronics.}
\end{thebibliography}

\section*{Acknowledgments}
The authors acknowledge the "COST Action MP1405 QSPACE", the European Union's Horizon 2020 and the EMPIR Participating States in the context of the project 17FUN01 BeCOMe, the John Templeton Foundation (Grant No. 43467) for financial support. The authors also acknowledge the Center for Macroscopic Quantum States (bigQ, DNRF412) of the Danish National Research Foundation and the Danish Council for Independent Research (Individual Postdoc and Sapere Aude 4184-00338B).

\section*{\textcolor{black}{Author contributions}}
\textcolor{black}{I.RB, P.T., I.P.D., M.G (responsible of INRIM quantum optics group), T.G., U.L.A. (responsible of DTU laboratories) planned the experiment. The experimental realization was achieved (supervised by the formers) by S.T.P. (leading role), E.L., C.S.J. and M.Z. Data analysis was mainly performed by  S.T.P., E.L. I.R.B., P.T., and M.Z, with some contribution from all authors. The manuscript was prepared with inputs by all the authors. They also had a fruitful systematic discussion on the progress of the work.}

\end{document}